\RequirePackage{fix-cm}
\documentclass[smallextended]{svjour3}       
\smartqed  
\usepackage{graphicx}
\begin{document}

\title{Integrating and validating urban simulation models}
\subtitle{}


\author{Juste Raimbault}


\institute{J. Raimbault \at
              Center for Advanced Spatial Analysis, University College London\\
              \email{j.raimbault@ucl.ac.uk}
}

\date{}

\maketitle

\begin{abstract}
\vspace{-1cm}
Urban systems are intrinsically complex, involving different dimensions and scales, and consequently various approaches and scientific disciplines. In that context, urban simulation models have been coined as essential for the construction of evidence-based and integrated urban sciences. This review and position paper synthesises previous work focused on coupling and integrating urban models on the one hand, and exploring and validating such simulation models on the other hand. These research directions are complementary basis for a research program towards the development of integrated urban theories, with some application perspectives to sustainable territorial planning.
\end{abstract}

\section{Introduction}

Cities, and more generally urban and territorial systems, are highly complex to understand, manage and plan. They imply imbricated scales, from the intra-urban microscopic scales with individual agents such as household or companies, through the intermediate mesoscopic scale of the urban area, to the macroscopic scale of systems of cities \cite{pumain2008socio}. Many different dimensions and viewpoints can be considered as an entry into urban issues, and several disciplines relate to emerging urban sciences \cite{batty2013new}, including urban economics, urban geography, planning and design. Complexity approaches to cities are therefore a relevant interdisciplinary take into urban systems \cite{batty2012origins}.

Furthermore, the question of sustainable cities and territories is crucial in terms of policy application of urban sciences, especially in the current context of global change and pressing need for sustainable transitions. Cities can be in that context as much an issue than a solution, and models with sufficient endogeneity can help to anticipate their dynamics. The integration of environmental dimensions within simulation models for territorial systems is in that context an important aspect, but the coupling between dimensions remains generally weak or to a certain level of exogeneity. For example, \cite{viguie2014downscaling} use an urban growth model to evaluate the local impact of climate change, but does not make associated aspects endogenous, such as energy price or the energy efficiency of the urban structure. Negative externalities such as congestion and pollution, directly linked to emissions, have a feedback for example on the location of activities. Similarly, numerous studies in ecology which establish the impact of anthropic habitat disturbances would particularly benefit from a coupling with urban growth models, for example for a better management of the interface between the city and nature within the new urban regimes that are urban mega-regions \cite{hall2006polycentric}. At the macroscopic scale, several couplings between models of systems of cities and ecological models or from environmental science can also be considered. For example issues related to the production, storage and distribution of energy are endogenous to territorial systems, and the dynamics of infrastructures and associated entities can be integrated within models for systems of cities \cite{pumain2018evolutionary}.

In that context, we propose a research program rooted into complexity science to foster the construction of bridges and interdisciplinary dialogues in urban science, in some sense the construction of integrated urban theories. The concept of integration can be understood through the complex systems roadmap \cite{bourgine2009french}. It combines horizontal integration (fundamental transversal questions at the intersection of different types of complex systems) with vertical integration (multiple levels coupled within multi-scale models). The integration between knowledge domains in the sense of \cite{raimbault2017applied}, (theoretical, empirical, modeling, data, methods and tools knowledge domains) is also an aspect of the expected integration, especially through the strong link between models, empirical data and methods, with their exploration and validation.

We indeed directly inherit from the evolutionary theory of cities developed by Denise Pumain \cite{pumain2017urban}, in particular regarding the role of simulation models and model validation methods - the OpenMOLE platform and associated methods \cite{reuillon2013openmole} were mainly developed in this frame. We assume thus that the construction of integrated models can not be achieved independently from their systematic exploration and validation. The project is thus built on three complementary axis: (i) the coupling of heterogenous urban models to integrate dimensions; (ii) the coupling of models at different scales for the vertical integration; and (iii) the development of new spatial simulation model validation methods, simultaneously applied to previous models.

The integration of urban models has a long history linked to the development of quantitative approaches to urban systems and operational urban models \cite{batty2021defining}. Land-use Transport Interaction models for example, which aim at simulating the impact of transportation infrastructure on different components of territories, generally include several sub-models such as transport models, cellular automatons for the evolution of land-use, or economic microsimulation models \cite{wegener2021land}. The framework proposed by \cite{batty2021new} implements spatial interaction modeling between population and employment on large-scale urban system, providing a basis for coupling with other dimensions and other types of models. More recently, the development of integrated models for smart cities has also put forward the relevance of model coupling in urban simulation. \cite{yamagata2013simulating} couple for example models for land-use and energy consumption. \cite{muvuna2019methodology} introduce a framework to integrate information from different components of the smart city.

The question of integrating models and dimensions has also been the focus of other disciplines not directly related to urban issues. Ecology has for example studied the interaction between society, ecosystems and resources \cite{bithell2009coupling}. The link between economics and ecology is a crucial issue regarding sustainability \cite{hofkes1996modelling}. Resource economics is a field in which model coupling can act as a medium for smoother interdisciplinary collaborations \cite{macleod2016model}. Within urban science, such research directions remain mostly to be explored, from diverse viewpoints including methodology and theory.

We describe each axis of our research program in the rest of this paper, mostly by developing a synthesis of previous research at the basis of this project. We finally discuss how the construction of such integrated models could be applied to the design of sustainable territorial planning policies.

\section{Horizontal integration: coupling urban models and dimensions}

\subsection{Model coupling}

An horizontal integration between complementary dimensions of urban systems can be achieved through the coupling of different urban models, taking into account different dimensions. The example of urban dynamics and environmental issues developed above can be investigated regarding several issues related to climate change, such as flooding \cite{ford2019multi} or heat waves \cite{lemonsu2015vulnerability}. The integration of economy and geography is also a question that remains to be mostly explored \cite{marchionni2004geographical}. More generally, the Sustainable Development Goals are declined into 17 goals and 169 targets which must be tackled in an integrative way \cite{stafford2017integration}.

Some previous work, building on the evolutionary theory of cities and at the inception of this project, focuses on the co-evolution of transportation networks and territories. It illustrates how such model coupling can be achieved. \cite{raimbault2019urban} couples a reaction-diffusion urban morphogenesis model with a multi-model of transportation network growth to yield a co-evolution model at the mesoscopic scale. At the scale of the system of cities, \cite{raimbault2021modeling} integrates an abstract urban network evolution model into an urban dynamics model for the population of cities. These models effectively achieve a strong coupling, as they capture circular causality between the transportation network and cities \cite{raimbault2020unveiling}. A refinement of the network growth heuristic for the macro model, described by \cite{raimbault2020hierarchy}, introduces an intermediate scale of geographical description, and is a first step towards multi-scale models.

Current work recently presented as \cite{raimbault2021estimating} and based on the methodology of \cite{raimbault2021building} focuses on coupling models to build open transport models from the bottom-up using open source components: the MATSim transport modeling framework \cite{horni2016multi} is for example integrated with the Quant spatial interaction model \cite{batty2021new} and a population microsimulation model \cite{lomax2017microsimulation}. The implementation of model coupling using the OpenMOLE workflow system \cite{passerat2017reproducible} illustrates the joint development and exploration of models. Expected policy results of this work integrate an additional dimension, through the implementation of density indicators in public transport which can be used as a proxy for potential contamination in the epidemiological context of COVID-19.

\subsection{Literature mapping}

Model coupling between disciplines is made simpler when a reflexive positioning is available, for example using scientometrics and literature mapping techniques. Therefore, \cite{raimbault2019exploration} provides a framework to jointly explore interdisciplinary citation and semantic networks, while \cite{raimbault2021empowering} develop open tools to explore scientific corpuses. 

Literature mapping and systematic reviews can also be used to identify and compare existing models with great detail, as for the case of land-use transport interactions: \cite{raimbault2021interdisciplinary} proposes a map of the interactions between disciplines involved in such modeling (from geography to urban economics, planning and physics) while \cite{raimbault2019meta} proceeds to a meta-analysis of model characteristics based on their scientific context. Such efforts can also be a basis towards more advanced model benchmarking experiments, as \cite{raimbault2020comparison} in the case of several urban morphogenesis models in terms of produced urban morphologies; these benchmark being in turn a step towards model integration or multi-modeling approaches.

\subsection{Open issues}

Several research questions remain rather open in the case of model coupling for urban systems, although some methodological and technical contributions could be imported from other disciplines to tackle them.

For instance, the definition of model coupling itself is not clear. A serial coupling of models (outputs of one becoming inputs of the other) can be qualified as weak or loose coupling \cite{clarke1998loose}, while the integration of feedback loops between model states at each time step, or the construction of a more general model including the two, can be seen as a strong coupling. Measures to quantify a degree of coupling also remain to be investigated.

Beyond the nature of the coupling, technical aspects must be investigated. When models have different time steps or even different time scales, implementations can not directly communicate. The DEVS framework implies that coupled components follow a same formalism and can be seamlessly integrated \cite{vangheluwe2000devs}. Methodological frameworks have been introduced in physics to couple asynchronous models across different time scales \cite{lockerby2015asynchronous}.

An other dimension is wether underlying epistemologies and ontologies are intrinsically compatible. In other words, the semantic contents of models may in some cases be in contradiction. Some disciplines (or even different schools of thought in the same discipline) may introduce conflicting assumptions. Making ontologies explicit, either using established ontology systems \cite{yang2019ontology}, or a less strict approach to ontology in the case of social systems \cite{livet2010ontology}, is a way to mitigate such issues.

A last difficulty worth mentioning is the computational complexity which may arise from model coupling. In the case of a strong integration in the sense of many feedback loop, this complexity may severely increase in comparison to the computational burden of models alone; notwithstanding the fact that coupling mechanically increases the number of parameters, making the integrating model difficult to explore due to the curse of dimensionality. Simplifying expectations on model outputs, working on patterns only as does the approach of Pattern Oriented Modeling \cite{grimm2005pattern}, may reduce uncertainties due to a higher complexity. Specific methods, such as inverse problems methods \cite{vogel2002computational}, allow targeted exploration of the parameter space and a reduction of complexity. One must still notice that different types of complexity are related \cite{raimbault2020relating}, and thus there is generally no straightforward way in terms of computation to take into account emergence in a model.

\section{Vertical integration: constructing multi-scale models}

The construction of multi-scale models is in itself a crucial aspect towards integrated models and theories. Following \cite{rozenblat2018conclusion}, systems of cities have reached multiple levels of articulation and interdependencies. This implies that their management and planning must necessarily be multi-scalar in order to take into account geographical particularities, while still ensuring a global consistence which is a condition for limited inequalities between territories.

Moreover, considerable methodological work is required to elaborate coupling methods between scales, as for example the hybrid modeling coupling agent-based models with differential equations for an epidemiological model \cite{banos2015coupling}. This allows determining the relevance of levels to be included for tackling a particular problem and avoid so-called ontological dead-ends (i.e. include appropriate levels of representation) \cite{roth2006reconstruction}. Such methodological investigation are also necessary to understand the nature of retroactions between scales to be included, and when these are necessary or not. A crucial aspect mostly neglected in the literature is to effectively achieve a strong coupling between scales, in the sense of including both upward and downward feedbacks.

Recent work paved the way towards such multi-scale strongly coupled models. \cite{raimbault2021strong} couples the urban dynamics model of \cite{raimbault2020indirect} at the scale of the system of cities with the urban morphogenesis model of \cite{raimbault2018calibration} at the scale of the urban area. Downward feedback is captured through policy parameters, as local development decisions generally react to the global integration of a given city, while upward feedback is taken into account by combining positive and negative externalities within the urban area to update its global performance and interaction parameters with other cities.

To understand the interplay between urban form at the microscopic scale, transportation network development and developer agents decisions at the mesoscopic scale, \cite{raimbault2021multiscale} proposes a stylised agent-based model achieving a strong coupling between the microscopic and mesoscopic scale.

These efforts remain at the stage of stylised and rather simple model, and the construction of data-driven and validated multi-scale models remains to be explored. Coupling between scales is also a way to couple between different urban dimensions, and both integration are not necessarily independent.

\section{Model exploration and validation}

\subsection{Model validation methods}

Working to integrate models and theories necessarily implies a better understanding of the modeling process itself, but also of stylised dynamics produced by simulation models, in the sense of patterns \cite{grimm2005pattern}.

In the case of geographical models for urban systems, such a knowledge has for example been developed within the \emph{Geodivercity} European Research Council project lead by Denise Pumain \cite{pumain2017urban}, which included to the conception of new model validation methods. These methods were elaborated specifically to tackle thematic geographical questions, but were applied to many contexts thereafter, witnessing a beneficial relationship both from the geography and computer science viewpoints \cite{raimbault2019methods}. More generally in social sciences, modeling and simulation driven by new practices including model coupling, the use of high-performance computing for model exploration, and open science practices, are tightly linked to the production of a new type of integrated knowledge \cite{banos2013pour}. This is in some sense an aspect of the computational shift in contemporary science coined by \cite{arthur2014complexity}.

Such model validation methods can be applied for example to assess the necessity and sufficiency of processes in a multi-modeling context, in the case of the Calibration Profile algorithm \cite{reuillon2015new}. Genetic algorithms distributed on a computation grid using OpenMOLE are a highly efficient tool for model calibration \cite{schmitt2015half}. The search for diversity in model outputs, obtained with the Pattern Space Exploration algorithm, can be used to explore the space of feasible configurations a model can produce, and possibly highlight unexpected behavior \cite{cherel2015beyond}.

Therefore, the development of specific methods and tools to improve the extraction of knowledge from simulation models, and an epistemological investigation on modeling practices, are crucial within this project. Methods for the exploration, sensitivity analysis, and validation, are essential for robust application of models, but also yield a better complementarity with other types of approaches since they can establish when modeling is not relevant anymore.

\subsection{Spatial sensitivity analysis}

Recent work has focused on the development of validation methods in the specific case of spatial simulation models. More particularly, a methodology coined as \emph{Spatial Sensitivity Analysis} has been introduced by \cite{raimbault2019space} in order to understand the effect of spatial initial conditions on model outcomes. This provides a generic way to disentangle effects which are intrinsic to model dynamics from contingent effects due to the geography. The generation of synthetic spatial configuration for territorial systems is required, and \cite{raimbault2019second} has investigated the generation of such data while controlling correlation patterns. \cite{raimbault2019generating} introduce and compare multiple generators for building configurations at the district scale. These methods are implemented into a single scala library, allowing an easier integration into the OpenMOLE software \cite{raimbault2020scala}.

\subsection{Further developments}

Different directions are worth exploring regarding the development of validation methods specific to spatial urban simulation models. Null models are rather rare in the field, in comparison to ecology for example which has used the Neutral Landscape Model for quite some time \cite{with1997use}. Bayesian calibration methods such as particle filters \cite{jiang2017new} are not widely used for urban models, although notable exceptions exists in domains close to engineering such as the study of traffic \cite{marinica2011particle}. Finally, some work on the nature and definition of validation itself would be required. Different standards are expected depending on disciplines and types of models, since for example a computational model will always be more difficult to trust than a tractable analytical model. Methods developed will depend on such definition and expected standards of robustness.

\section{Discussion: towards evidence-based multi-scalar sustainable territorial planning}

The strong complementarity of the three axis of the research program, elaborated on previous and current work described above, stems naturally from the intrinsic link between vertical and horizontal integrations, and from the integration of knowledge domains as model validation methods are elaborated jointly as models are constructed and explored. A fully open issue is the transfer of integrated models towards policy applications.

Models have multiple functions, for which a precise typology has been introduced by \cite{varenne2018theories}. Regarding the theoretical and quantitative geography \cite{cuyala2014analyse} side of our proposal, in terms of theoretical contribution, models are instruments of knowledge production and contribute to the construction of theories. Transferring models towards policy applications requires to diversify their functions, and the end users. In a perspectivist view of knowledge production \cite{giere2010scientific}, models can not be dissociated from the question they are designed to answer and from the entity formulating it. Model application implies thus an upstream design, or redesign, of models such that planners, policy makers, stakeholders, and the public - any relevant user indeed - are blent in the modelling process from the beginning.

Concrete research directions have also to be investigated for a possible application of models. A first research direction lies in the modeling of policy and governance itself. The management of sustainability implies according to \cite{etzion2018management} new governance structures. Their modeling is still not well integrated within spatially explicit models. \cite{le2015modeling} has proposed to integrate transportation network governance into a land-use transport model to yield a co-evolution model. \cite{raimbault2020coevolution} uses game-theory to simulate transportation network investments at the scale of the system of cities.

One other crucial research axis towards the application of integrated models to policies is the construction of harmonised multi-dimensional databases. Such databases must (i) consider consistent geographical entities in time, in the sense of the dynamical ontology proposed by \cite{bretagnolle2009villes}; (ii) integrate multiple dimensions to allow coupling urban dynamics with socio-economic, environmental, sustainability and global change issues; (iii) be open, accessible, documented and reproducible. Technical aspects of data integration will also have to be carefully taken into account. In that regard, contributions from digital twins modeling (a model simulating a system in real time and having potential feedback on its functioning) applied to urban systems \cite{batty2018digital}, or tools and methods to handle information elaborated in the context of smart cities \cite{shahrour2017smart}, are important contributions.

Finally, models must be constructed jointly with the precision of their application context to ensure their potential applicability. Obstacles to applicability can raise on diverse dimensions. Regarding the aforementioned model end users, many different paths can be taken to ensure a co-construction, such as companion modeling \cite{drogoul2015agent} in which local stakeholders participate to the construction and exploration of the model. Issues related to real-time model visualisation are also an important technical problem to take into account for model applications \cite{milton2019accelerating}. Scientific mediation initiatives are an other way to diffuse the results of modeling experiments into policy applications.

\section{Conclusion}

The diversity and complexity of urban systems requires a plurality of viewpoints \cite{pumain2020conclusion}, which call for integration between dimensions and scales. We have synthesised in this paper a current stream of research focusing around three axis, namely (i) horizontal integration through model coupling; (ii) vertical integration through the construction of multi-scale urban models; and (iii) development of model validation and exploration methods. These strongly complementary axis form the basis of a long term research program, with a goal of application to sustainable territorial policies.


\end{document}